\documentclass[letter,twocolumn]{jpsj3}
\usepackage{color}

\renewcommand{\a}{\alpha}
\renewcommand{\b}{\beta}

\newcommand{\bea}{\begin{eqnarray}}
\newcommand{\eea}{\end{eqnarray}}
\newcommand{\f}[2]{\frac{#1}{#2}}
\newcommand{\eq}{&=&}
\newcommand{\nn}{\nonumber \\ }
\newcommand{\ve}{\varepsilon}

\newcommand{\area}{\int_{-\infty}^\infty }

\newcommand{\p}{\partial}
\newcommand{\pp}[2]{\f{\p #1}{\p #2}}

\newcommand{\sref}[1]{Eq. (\ref{#1})}

\allowdisplaybreaks[1]

\title{Replica Analysis for Portfolio Optimization \\
with Single-Factor Model
}

\author{
Takashi Shinzato\thanks{shinzato@eng.tamagawa.ac.jp}
}
\inst{%2-1 Naka, Kunitachi, Tokyo 186-8601, Japan
Department of Management Science, College of Engineering, 
Tamagawa University, \\ Machida, Tokyo, 1948610, Japan\\
\smallskip
{\rm (Received April 5, 2017; accepted *** 1, 2017)}
} %\\

\abst{
{
In this paper, we use replica analysis to
investigate the influence of correlation 
among the return rates of assets 
on 
the solution of the portfolio optimization problem.
We consider the behavior of the optimal solution for the case where 
the return rate is described with a single-factor model and compare the findings obtained from our proposed methods with correlated return rates with 
those obtained with 
independent return rates. We then analytically assess 
the increase in the investment risk when correlation is included. 
Furthermore, we also compare our approach with 
analytical procedures for minimizing the investment risk from operations research. 
}
}

\kword{mean-variance model, single-factor model, investment risk, investment concentration, replica analysis}
\begin{document}
\maketitle

{
In recent decades, 
investment strategies for the portfolio optimization problem have been considered extensively using a combination of analytical approaches from different research fields,
including
{econophysics and statistical mechanical informatics} \cite{Ciliberti1,
Ciliberti2,
Kondor,
Caccioli,
Pafka,
Shinzato-2015-PLOS7,
Shinzato-2015-PLOS8,
Shinzato-2016-PRE11,
VH,
Shinzato-2016-PRE12,
Shinzato-2017-JSTAT2,
Shinzato-2011-IEICE,
Kondor-2016-with-no-short-selling,
Shinzato-ve-fixed2016,
Shinzato-Pythagorean2017}.
Recently, the mean-variance model, which is one of the most popular portfolio optimization 
{problems}, 
has been the subject of renewed interest in a variety of cross-disciplinary studies \cite{Pafka,
Shinzato-2015-PLOS7,
Shinzato-2015-PLOS8,
Shinzato-2016-PRE11,
VH,
Shinzato-2016-PRE12,
Shinzato-2017-JSTAT2,
Shinzato-2011-IEICE,
Kondor-2016-with-no-short-selling,
Shinzato-ve-fixed2016,
Shinzato-Pythagorean2017}.
In particular, 
the objective function for the investment risk in the mean-variance model
is mathematically similar 
to the Hamiltonian of the Hopfield model,
which has been widely used in studies on the associative memory problem, as both objective functions
are described by using the 
quadratic form with respect to 
thermodynamic variables, and Hebb's rule is related to 
the variance-covariance matrix of the return rate {\cite{Shinzato-2015-PLOS7}.}
{The} optimal portfolio which minimizes the investment risk is also interpreted as
corresponding to the ground state in the spin glass model,
and consequently, several previous studies have 
applied techniques that were developed in spin glass theory 
such as replica analysis, belief propagation, and random matrix theory 
to investigate the optimal portfolio.
}

{Although in \cite{Pafka,
Shinzato-2015-PLOS7,
Shinzato-2015-PLOS8,
Shinzato-2016-PRE11,
VH,
Shinzato-2016-PRE12,
Shinzato-2017-JSTAT2,
Shinzato-2011-IEICE,
Kondor-2016-with-no-short-selling,
Shinzato-ve-fixed2016,
Shinzato-Pythagorean2017} it is usually assumed that the 
return rates are independent,
the return rates of assets in actual investment portfolios may be correlated,
meaning that the models developed in these studies may underestimate the risk of loss (negative return rates) and should be used with caution.
To analyze the portfolio 
optimization problem analytically with correlated return rates,
we need to utilize and extend existing methods 
from a variety of fields.
As a first step for characterizing the correlation among return rates,
we consider a single-factor model that is widely used 
in mathematical finance and 
discuss whether the 
optimal portfolio which minimizes the investment risk 
with budget constraints is affected by correlation among 
the return rates 
using replica analysis.
}

\if 0
\begin{table}
\caption{List of options for paper types.}
\label{t1}
\begin{center}
\begin{tabular}{ll}
\hline
\multicolumn{1}{c}{Option} & \multicolumn{1}{c}{Paper type} \\
\hline
\verb|ip| & Invited Review Papers \\
\verb|st| & Special Topics \\
\verb|letter| & Letters \\
\verb|fp| & Full Papers \\
\verb|shortnote| & Short Notes \\
\verb|comment| & Comments \\
\verb|addenda| & Addenda \\
\verb|errata| & Errata \\
\hline
\end{tabular}
\end{center}
\end{table}
\fi

{Following previous work, 
we begin by considering the situation
where rational investors invest 
into $N$ assets over $p$ periods in a 
steady investment market 
with no short-selling. The {portfolio of asset} $i(=1,2,\cdots,N)$ {is} 
denoted by $w_i\in{\bf R}$, and 
$\vec{w}=(w_1,w_2,\cdots,w_N)^{\rm T}\in{\bf R}^N$ is
the entire portfolio, where T denotes {its} transpose.
Since there is no 
short-selling, we note that 
$w_i$ is not always positive. Furthermore, $\bar{x}_{i\mu}$
indicates the return rate of asset $i$
{at} period $\mu(=1,2,\cdots,p)$ and its expectation is $E[\bar{x}_{i\mu}]$.
Then, in {investing periods}, the investment risk of 
portfolio $\vec{w}$, ${\cal H}(\vec{w}|X)$, is defined as follows:
\bea
\label{eq1}
{\cal H}(\vec{w}|X)\eq\f{1}{2N}
\sum_{\mu=1}^p
\left(\sum_{i=1}^N\bar{x}_{i\mu}w_i
-
\sum_{i=1}^NE[\bar{x}_{i\mu}]w_i
\right)^2\nn
\eq\f{1}{2}\vec{w}^{\rm T}J\vec{w},
\eea
where $x_{i\mu}=\bar{x}_{i\mu}-E[\bar{x}_{i\mu}]$ is 
the modified return rate and the return rate matrix $X=\left\{\f{x_{i\mu}}{\sqrt{N}}\right\}\in{\bf R}^{N\times p}$
is defined using the modified return rates, and 
entry $i,j$ of 
the variance-covariance {(or Wishart)} matrix 
$J=\left\{J_{ij}\right\}(=XX^{\rm T})\in{\bf R}^{N\times N}$ is 
$J_{ij}=\f{1}{N}\sum_{\mu=1}^px_{i\mu}x_{j\mu}=(XX^{\rm T})_{ij}$. Here, the budget constraint
\bea
\label{eq2}
\sum_{i=1}^Nw_i\eq N
\eea
is used. From this, we need to 
determine the optimal portfolio which minimizes
the investment risk ${\cal H}(\vec{w}|X)$ in \sref{eq1} 
from the set of portfolios that satisfy the budget constraint in \sref{eq2}.
With respect to the optimal portfolio $\vec{w}^*=\arg\mathop{\min}_{\vec{w}\in{\cal W}}{\cal H}(\vec{w}|X)$,
determining analytically the minimal investment risk per asset $\ve=\f{1}{N}{\cal H}(\vec{w}^*|X)$
and its investment concentration $q_w=\f{1}{N}(\vec{w}^{*})^{\rm T}
\vec{w}^{*}$ 
is one of the most active {issues being researched}
for the portfolio optimization problem, 
and a variety of cross-disciplinary approaches have been developed.
Here, 
${\cal W}=\left\{\vec{w}\in{\bf R}
\left|
\sum_{i=1}^Nw_i=N
\right.
\right\}$ is the feasible subset of portfolios satisfying \sref{eq2}. 
Our previous work\cite{Shinzato-2015-PLOS7}
discussed the case where 
$x_{i\mu}$ is independently and identically distributed with 
mean 0 and variance 1,
and the minimal investment risk per asset $\ve$
and {its} investment concentration $q_w$ were determined 
as follows:
\bea
\label{eq3}
\ve\eq\f{\a-1}{2},\\
\label{eq4}
q_w\eq\f{\a}{\a-1}.
\eea
For the case where $x_{i\mu}$ is independently distributed with
mean 0 and variance $v_i$, that is,
the variance of each asset is distinct, 
the minimal investment risk per asset $\ve$
and its investment concentration $q_w$ were {also} determined 
as follows:
\bea
\label{eq5-1}
\ve\eq\f{\a-1}{2\left\langle v^{-1}\right\rangle},\\
\label{eq6}
q_w\eq\f{\left\langle v^{-2}\right\rangle}{\left\langle v^{-1}\right\rangle^2}
+\f{1}{\a-1},
\eea
where $\a=p/N\sim O(1)$ \cite{Shinzato-2016-PRE12}.
In order to determine uniquely the optimal portfolio $\vec{w}^{*}$,
the squared matrix $J$ should be 
regularized, and then 
the above-mentioned results hold for $\a>1$. 
Similarly,
in the present work, we
assume for $\a>1$, the optimal portfolio is uniquely determined.
Moreover,
the notation 
$\left\langle g(v)\right\rangle=\lim_{N\to\infty}\f{1}{N}\sum_{i=1}^Ng(v_i)$
is used.
}

{Namely, in}
{previous work,
the return rates $x_{i\mu}$ were assumed to be 
independently and identically distributed with mean 0 and variance 1, or independently (but not identically)
distributed with mean 0 and variance $v_i$.
However,
the return rates of assets in many practical situations
are correlated,
and the findings in previous work which 
assumed independent rates 
may be unsuitable for practical applications, as they will underestimate the investment risk. 
Thus, 
as a first step for 
characterizing the correlations among return rates, 
we should analyze 
the minimal investment risk per asset $\ve$ and 
its investment concentration $q_w$ 
for the portfolio minimizing the investment risk 
for the case where 
the return rate of each asset is determined with a 
single-factor model. Here, 
using a single-factor model, the return rate 
$x_{i\mu}$ is defined as follows:
\bea
\label{eq7-1}
x_{i\mu}\eq \f{1}{\sqrt{N}}b_if_\mu+y_{i\mu},
\eea
where $\f{1}{\sqrt{N}}$ is
the scaling parameter which can be adjusted 
to simplify the analytical results.
Moreover,
$f_\mu$ is the macroeconomic indicator at period 
$\mu$ (the probability of
$f_\mu$ is already known and its mean is assumed to be 0, and we
do not require the indicator to be normally distributed),
and $b_i$ denotes the level of influence of the
macroeconomic indicator $f_{\mu}$ on asset $i$.
(Hereafter we call this the factor loading. The probability of $b_i$ is also assumed to be known and
and does not need to be normally distributed.) Further,
the (independent) return rate $y_{i\mu}$ is independent of the other return rates and 
is not correlated with macroeconomic indicator $f_\mu$
and factor loading $b_i$, and 
the mean and the variance are 0 and $v_i${, respectively.}
That is,
$x_{i\mu}$ in \sref{eq7-1} is regarded as 
a linear regression equation with noise term $y_{i\mu}$. 
In general, since 
macroeconomic indicators 
may include temporal trends, we
do not assume independence {among} macroeconomic indicators.
Similarly, there may exist 
correlation among factor loadings, 
and the assumption of independence among factor loadings is not required
in this work.}

\if 0
\begin{figure}
%\includegraphics{fig01.eps}
\caption{You can embed figures using the \texttt{\textbackslash includegraphics} command. EPS is the only format that can be embedded. Basically, figures should appear where they are cited in the text. You do not need to separate figures from the main text when you use \LaTeX\ for preparing your manuscript.}
\label{f1}
\end{figure}
\fi

{Let us reformulate the above optimization problem 
in the framework of statistical mechanical informatics
and analyze the minimal investment risk per asset $\ve$ and its investment concentration 
$q_w$
using replica analysis. First, from the framework of 
statistical mechanical informatics, 
the partition function $Z(\b,X)$ at inverse temperature 
$\b(>0)$ is defined as follows:
\bea
Z(\b,X)\eq\int_{\vec{w}\in{\cal W}}
d\vec{w}
e^{-\b{\cal H}(\vec{w}|X)}.
\eea
From this, we can determine 
the average of 
the logarithm of the partition function 
per asset as follows:
\bea
\label{eq5}
\phi\eq\lim_{N\to\infty}
\f{1}{N}E\left[\log Z(\b,X)\right]\nn
\eq\lim_{N\to\infty}
\lim_{n\to0}\f{1}{N}\pp{}{n}\log E\left[Z^n(\b,X)\right].
\eea
From the formula
\bea
\label{eq10}
\ve\eq-\lim_{\b\to\infty}
\pp{\phi}{\b},
\eea
we can evaluate the minimal investment risk per asset analytically, where
the notation $E[g(X)]$ means the expectation of $g(X)$ with respect to the return rate. From replica analysis,
\bea
\label{eq7}
\phi\eq\mathop{\rm Extr}_\Theta
\left\{
-k-hm+\f{1}{2}(\chi_w+q_w)(\tilde{\chi}_w-\tilde{q}_w)
\right.\nn
&&+\f{1}{2}q_w\tilde{q}_w+\f{1}{2}(\chi_s+q_s)(\tilde{\chi}_s-\tilde{q}_s)+\f{1}{2}q_s\tilde{q}_s
\nn
&&-\f{\a}{2}\log(1+\b\chi_s)-\f{\a\b (q_s+Fm^2)}{2(1+\b\chi_s)}
\nn
&&
-\f{1}{2}\left
\langle
\log
(\tilde{\chi}_w+v\tilde{\chi}_s)
\right\rangle
\nn
&&
\left.
+
\f{1}{2}\left
\langle
\f{\tilde{q}_w+v\tilde{q}_s+(k+bh)^2}{\tilde{\chi}_w+v\tilde{\chi}_s}
\right\rangle
\right\},
\eea
is obtained, where 
$\mathop{\rm Extr}_rg(r)$ is the extremum of 
$g(r)$ with respect to the parameter 
$r$ and $\Theta=\left\{k,m,h,\chi_w,q_w,\tilde{\chi}_w,\tilde{q}_w
,\chi_s,q_s,\tilde{\chi}_s,\tilde{q}_s
\right\}$ represents the set of order parameters,
\bea
\label{eq12}
F\eq\lim_{p\to\infty}\f{1}{p}
\sum_{\mu=1}^pf_\mu^2,\\
\label{eq13}
\left\langle
g(b,v)
\right\rangle
\eq\lim_{N\to\infty}
\f{1}{N}\sum_{i=1}^Ng(b_i,v_i),
\eea
and $\a=p/N\sim O(1)$ (see Appendix for further details).
Note that since $F$ in \sref{eq12} is
the average of the square of the macroeconomic indicators,
we can determine $F$ easily, regardless of the presence or absence of correlation among the macroeconomic 
indicators. In addition,
from \sref{eq13}, it is also 
easy to assess $\left\langle g(b,v)\right\rangle$
regardless of the presence or absence of correlation among factor loadings.
From the above,
\bea
\chi_w\eq\f{\left\langle v^{-1}\right\rangle}{\b(\a-1)},\\
\label{eq15}
q_w\eq\f{1}{\a-1}\left(1+Fmm_1\left\langle v^{-1}
\right\rangle\right)+C
,\\
\chi_s\eq\f{1}{\b(\a-1)},\\
q_s\eq
\f{1}{\left\langle v^{-1}\right\rangle
}
+F^2m^2 V_1\left\langle v^{-1}\right\rangle
\nn
&&+\f{1}{\a-1}\left(
\f{1}{\left\langle v^{-1}\right\rangle}+
Fmm_1
\right),
\eea
{are determined,} where {
\bea
\label{eq18}
m\eq\f{m_1}{1+FV_1
\left\langle v^{-1}\right\rangle
}.
\eea
Furthermore, $m_1=\f{\left\langle v^{-1}b\right\rangle}
{\left\langle v^{-1}\right\rangle}$, 
$V_1=\f{\left\langle v^{-1}b^2\right\rangle}
{\left\langle v^{-1}\right\rangle}
-\left(\f{\left\langle v^{-1}b\right\rangle}
{\left\langle v^{-1}\right\rangle}\right)^2$, 
$m_2=\f{\left\langle v^{-2}b\right\rangle}
{\left\langle v^{-2}\right\rangle}
$, $V_2=\f{\left\langle v^{-2}b^2\right\rangle}
{\left\langle v^{-2}\right\rangle}
-\left(\f{\left\langle v^{-2}b\right\rangle}
{\left\langle v^{-2}\right\rangle}\right)^2
$, and $C=F^2m^2V_2\left\langle v^{-2}
\right\rangle+\f{\left\langle v^{-2}\right\rangle}{\left\langle v^{-1}\right\rangle^2}
\left\{
1+Fm(m_1-m_2)\left\langle
v^{-1}
\right\rangle
\right\}^2$.
}
\if 0
\bea
\label{eq18}
m\eq\f{m_1}{1+FV_1
\left\langle v^{-1}\right\rangle
},\\
m_1\eq\f{\left\langle v^{-1}b\right\rangle}
{\left\langle v^{-1}\right\rangle},\\
V_1\eq\f{\left\langle v^{-1}b^2\right\rangle}
{\left\langle v^{-1}\right\rangle}
-\left(\f{\left\langle v^{-1}b\right\rangle}
{\left\langle v^{-1}\right\rangle}\right)^2,\\
m_2\eq\f{\left\langle v^{-2}b\right\rangle}
{\left\langle v^{-2}\right\rangle},\\
V_2\eq\f{\left\langle v^{-2}b^2\right\rangle}
{\left\langle v^{-2}\right\rangle}
-\left(\f{\left\langle v^{-2}b\right\rangle}
{\left\langle v^{-2}\right\rangle}\right)^2,\\
C\eq F^2m^2V_2\left\langle v^{-2}
\right\rangle\nn
&&
+\f{\left\langle v^{-2}\right\rangle}{\left\langle v^{-1}\right\rangle^2}
\left\{
1+Fm(m_1-m_2)\left\langle
v^{-1}
\right\rangle
\right\}^2.\quad
\eea
\fi

From this, the minimal investment risk per asset 
$\ve$ is derived using \sref{eq10}, 
$\ve=
-\lim_{\b\to\infty}\pp{\phi}{\b}=
\lim_{\b\to\infty}\left\{\f{\a\chi_s}{2(1+\b\chi_s)}+
\f{\a(q_s+Fm^2)}{2(1+\b\chi_s)^2}
\right\}$ as follows:
\bea
\label{eq24}
\ve\eq\f{\a-1}{2
\left\langle v^{-1}\right\rangle}
+\f{\a-1}{2}Fmm_1.
\eea
We note that $mm_1\ge0$ is determined from \sref{eq18}, so that 
this {findings} 
is not smaller than the one obtained in our previous work, {(or see \sref{eq5-1}).}
}

{We now
consider whether the models obtained in the present work include 
the results obtained in previous work 
as special cases. First, from the assumption of independent return rates that are not influenced by macroeconomic indicators, 
that is, when $b_i=0$, $m=m_1=m_2=0$, and $V_1=V_2=0$, 
Eqs. (\ref{eq15}) and (\ref{eq24}) become
\bea
\label{eq28}
\ve\eq\f{\a-1}{2{\left\langle v^{-1}\right\rangle}},\\
\label{eq29}
q_w\eq\f{1}{\a-1}+
\f{\left\langle v^{-2}\right\rangle}{\left\langle v^{-1}\right\rangle^2},
\eea
where $C=\f{\left\langle v^{-2}\right\rangle}{\left\langle v^{-1}\right\rangle^2}$.
These equations are consistent with the results of previous work (Eqs. (\ref{eq5-1}) and (\ref{eq6})).
Further, 
the second term in \sref{eq24}, $\f{\a-1}{2}Fmm_1$, quantifies the influence from common factors $f_1,f_2,\cdots,f_p$ in a single-factor model.
Note that $Fmm_1$ is a monotonic nondecreasing function 
with respect to $F$, $\lim_{F\to0}Fmm_1=0$ and 
$\lim_{F\to\infty}Fmm_1=\f{m_1^2}{V_1\left\langle v^{-1}\right\rangle}$. In addition, 
although it is a superfluous consideration, 
for the case where
the variance of the return rate of each asset is unique, that is,
when $v_i=1$,
by substituting $\left\langle v^{-1}\right\rangle=\left\langle v^{-2}\right\rangle=1$
into Eqs. (\ref{eq28}) and (\ref{eq29}),
Eqs. (\ref{eq3}) and (\ref{eq4}) are obtained.
Since our results 
include 
the findings obtained in previous work 
as special cases, it is confirmed that 
our model is a natural extension which can handle 
the case of correlated return rates.
}

{
Finally, we 
compare the minimum expected investment risk 
which is 
obtained with an analytical procedure that is well known in operations research. 
First, from the 
portfolio which minimizes the expected investment risk $E[{\cal H}(\vec{w}|X)]$, 
that is, $\vec{w}^{\rm OR}=\arg\mathop{\min}_{\vec{w}\in{\cal W}}
E[{\cal H}(\vec{w}|X)]$, the minimum expected investment risk per asset $\ve^{\rm OR}=\lim_{N\to\infty}\f{1}{N}E[{\cal H}(\vec{w}^{\rm OR}|X)]$ can
easily be obtained as follows:
\bea
\ve^{\rm OR}\eq\f{\a}{2
\left\langle v^{-1}\right\rangle}
+\f{\a}{2}Fmm_1.
\eea
From this, the 
opportunity loss 
$\kappa=\f{\ve^{\rm OR}}{\ve}$ is computed as follows:
\bea
\kappa\eq\f{\a}{\a-1}.
\eea
Using a similar argument as in our previous work \cite{Shinzato-Pythagorean2017}, 
we note that although the opportunity loss $\kappa$ depends on the period ratio $\a$, 
it does not depend on the statistical properties of $v_i,b_i,f_\mu$. 
Moreover, from the investment concentration of the 
portfolio $\vec{w}^{\rm OR}$ which is
derived analytically using a procedure from operations research, 
{$q_w^{\rm OR}=\f{1}{N}(\vec{w}^{\rm OR})^{\rm T}\vec{w}^{\rm OR}$} is calculated as follows:
\bea
q_w^{\rm OR}\eq F^2m^2V_2\left\langle v^{-2}
\right\rangle\nn
&&
+\f{\left\langle v^{-2}\right\rangle}{\left\langle v^{-1}\right\rangle^2}
\left\{
1+Fm(m_1-m_2)\left\langle
v^{-1}
\right\rangle
\right\}^2,\quad
\eea
where it is found that $q_w^{\rm OR}$ corresponds to 
the last term $C$ in the investment concentration $q_w$ of 
the optimal portfolio $\vec{w}^{*}$ in \sref{eq15}. 
As noted in \cite{Shinzato-2015-PLOS7}, 
since rational investors 
prefer to invest in assets whose risks are 
comparatively low, in the investing periods when $\a$ is close to 1 the risks of the assets vary greatly and 
the investment concentration of the optimal portfolio increases. In contrast, 
when $\a$ is sufficiently large, 
the risks of the assets are almost indistinguishable in terms of the return rates, and 
rational investors will invest equally across all assets;
therefore, the investment concentration will tend to be low. This behavior is
reflected in 
our proposed approach; however,
the investment concentration of the 
portfolio $\vec{w}^{\rm OR}$ derived with the approach from 
operations research, $q_w^{\rm OR}$, is 
always constant with period ratio $\a$, and 
this is inconsistent with the optimal investment behavior of the rational investors. 
We have also verified that the analysis of the annealed disordered systems (related to the ordinary operations research approach) 
is distinct from
the analysis of quenched disordered systems (the analytical procedure based on our proposed method which corresponds to the analysis of the optimal investment strategy).
}

{
In the present work, we 
have 
analyzed the minimization of the investment risk with 
budget constraints 
for the case of 
correlated return rates
using a cross-disciplinary replica analysis approach from 
econophysics and statistical mechanical informatics. As there are many different types of dependence among the return rates of assets in an actual investment market, 
we used the single-factor model, as it is one of
the most fundamental models for correlation among return rates in mathematical finance.
{We discussed whether the correlation among return rates characterized by a single-factor model would influence the optimal solution.}
Further,
we compared our approach with 
the findings obtained from previous work and 
verified the effectiveness of the methodology proposed here for determining analytically and explicitly the investment risk in the presence of correlated assets.
}

{In actual investment markets, there are a 
myriad of macroeconomic indicators, and thus, in future work we will try to adapt the techniques developed for the associative memory problem \cite{Amit1,Amit2,Amit3},
to apply them to the optimization problem when 
the number of factors is $O(1)$ and 
$O(N)$. 
This paper discussed the investment risk minimization problem with budget constraints 
only, while 
the investment risk minimization problem in practice involves 
several constraints, for instance, the expected return and investment concentration {constraints}, 
and we need to analyze how these additional constraints influence 
the optimal solution for minimizing the investment risk 
with and without correlated
return rates.
}

%\begin{acknowledgment}
{The author thanks K.
Kobayashi and D. Tada for their valuable discussions. This work was supported partly by
Grant-in-Aid No. 15K20999; the President Project for Young
Scientists at Akita Prefectural University; Research Project No.
50 of the National Institute of Informatics, Japan; Research
Project No. 5 of the Japan Institute of Life Insurance; the Institute of Economic Research Foundation at
Kyoto University; Research Project No. 1414 of the Zengin
Foundation for Studies in Economics and Finance; Research
Project No. 2068 of the Institute of Statistical Mathematics;
Research Project No. 2 of the Kampo Foundation; and the Mitsubishi UFJ Trust Scholarship Foundation.
}
%\end{acknowledgment}

\appendix
\section{}
{
In this appendix, 
we derive 
$\phi$ using replica analysis.
Following the discussion in our previous work \cite{Shinzato-Pythagorean2017},
$E[Z^n(\b,X)],(n\in{\bf Z})$ is described as follows:
\bea
&&E\left[Z^n(\b,X)\right]\nn
\eq\mathop{\rm Extr}_{\vec{k}}
\f{1}{(2\pi)^{\f{Nn}{2}+pn}}
\area \prod_ad\vec{w}_a
d\vec{u}_a
d\vec{z}_a\nn
&&E\left[
\exp\left(
-\f{\b}{2}\sum_{\mu,a}z_{\mu a}^2
+\sum_ak_a
\left(\sum_iw_{ia}-N\right)
\right.
\right.\nn
&&
\left.\left.
+i
\sum_{\mu,a}u_{\mu a}
\left(z_{\mu a}-\f{1}{\sqrt{N}}\sum_{i}w_{ia}x_{i\mu}\right)
\right)\right],
\eea
where $\prod_a$ displays $\prod_{a=1}^n$, $\sum_i$ is $\sum_{i=1}^N$, 
$\sum_\mu$ means $\sum_{\mu=1}^p$, 
and $\sum_a$ represents $\sum_{a=1}^n$. Moreover,
$\vec{w}_a=(w_{1a},w_{2a},\cdots,w_{Na})^{\rm T}\in{\bf R}^N$, 
$\vec{u}_a=(u_{1a},u_{2a},\cdots,u_{pa})^{\rm T}\in{\bf R}^p$, 
$\vec{z}_a=(z_{1a},z_{2a},\cdots,z_{pa})^{\rm T}\in{\bf R}^p,(a=1,\cdots,n)$,
$\vec{k}=(k_1,k_2,\cdots,k_n)^{\rm T}\in{\bf R}^n$, and 
$k_a$ is the auxiliary parameter related to the budget constraint in \sref{eq2}. Next,
the order parameters are defined by
\bea
m_a\eq\f{1}{N}\sum_{i=1}^Nb_iw_{ia},\\
q_{wab}\eq\f{1}{N}\sum_{i=1}^Nw_{ia}w_{ib},\\
q_{sab}\eq\f{1}{N}\sum_{i=1}^Nv_iw_{ia}w_{ib},
\eea
and $h_a,\tilde{q}_{wab},\tilde{q}_{sab}$ are the conjugate parameters, where $a,b=1,2,\cdots,n$. 
From the ansatz of the replica symmetry solution, which comprises $k_a=k$, $m_a=m$, $h_a=h$, 
$q_{waa}=\chi_w+q_w$, $q_{wab}=q_w,(a\ne b)$, 
$\tilde{q}_{waa}=\tilde{\chi}_w-\tilde{q}_w$, $\tilde{q}_{wab}=-\tilde{q}_w,(a\ne b)$, 
$q_{saa}=\chi_s+q_s$, $q_{sab}=q_s,(a\ne b)$, 
and $\tilde{q}_{saa}=\tilde{\chi}_s-\tilde{q}_s$, $\tilde{q}_{sab}=-\tilde{q}_s,(a\ne b)$, the following is obtained:
\bea
&&
\lim_{N\to\infty}\f{1}{N}\log 
E\left[Z^n(\b,X)\right]\nn
\eq
\mathop{\rm Extr}_\Theta
\left\{
-nk-nhm+\f{n}{2}(\chi_w+q_w)(\tilde{\chi}_w-\tilde{q}_w)
\right.\nn
&&
-\f{n(n-1)}{2}q_w\tilde{q}_w
+\f{n}{2}(\chi_s+q_s)(\tilde{\chi}_s-\tilde{q}_s)\nn
&&
-\f{n(n-1)}{2}q_s\tilde{q}_s
-\f{\a(n-1)}{2}\log(1+\b\chi_s)\nn
&&-\f{\a}{2}\log(1+\b\chi_s+n\b q_s)
-\f{n\a\b Fm^2}{2(1+\b\chi_s+n\b q_s)}
\nn
&&-\f{n-1}{2}
\left\langle
\log(\tilde{\chi}_w+v\tilde{\chi}_s)
\right\rangle
\nn
&&-\f{1}{2}
\left\langle
\log(\tilde{\chi}_w+v\tilde{\chi}_s-n\tilde{q}_w-nv\tilde{q}_s)
\right\rangle
\nn
&&
\left.
+\f{n}{2}
\left\langle
\f{(k+bh)^2}{\tilde{\chi}_w+v\tilde{\chi}_s-n\tilde{q}_w-nv\tilde{q}_s}
\right\rangle
\right\}. 
\eea
We substitute this result into \sref{eq5} and obtain 
\sref{eq7}.
}
%\bibliography{shinzato-bib}
%\end{document}

\end{document}